\documentclass{revtex4-2}
\usepackage{amsmath}
\usepackage{graphicx}
\usepackage{color}
\begin{document}

\title{An amended Ehrenfest theorem for the Gross-Pitaevskii equation in
one- and two-dimensional potential boxes }
\author{Hidetsugu Sakaguchi$^{1}$ and Boris A. Malomed$^{2,3}$}
\address{$^{1}$Department of Applied Science for Electronics and Materials, Interdisciplinary Graduate School of
Engineering Sciences, Kyushu University, Kasuga, Fukuoka 816-8580, Japan}
\address{$^{2}$Department of Physical Electronics, School of Electrical Engineering,
Faculty of Engineering, and Center for Light-Matter Interaction, Tel Aviv
University, P.O. Box 39040 Tel Aviv, Israel}
\address{$^{3}$Instituto de Alta Investigaci\'{o}n, Universidad de Tarapac\'{a}, Casilla 7D,
Arica, Chile}

\begin{abstract}
It is known that the usual form of the Ehrenfest theorem (ET), which couples
the motion of the center of mass (COM) of the one-dimensional (1D) wave
function to the respective classical equation of motion, is not valid in the
case of the potential box, confined by the zero boundary conditions. A
modified form of the ET was proposed for this case, which includes an
effective force originating from the interaction of the 1D quantum particle
with the box' edges. In this work, we derive an amended ET for the
Gross-Pitaevskii equation (GPE), which includes the cubic nonlinear term, as
well as for the 2D square-shaped potential box. In the latter case, we
derive an amended COM equation of motion with an effective force exerted by
edges of the rectangular box, while the nonlinear term makes no direct
contribution to the 1D and 2D versions of the ET. Nonetheless, the
nonlinearity affects the amended ET through the edge-generated force. As a
result, the nonlinearity of the underlying GPE can make the COM motion in
the potential box irregular. The validity of the amended ET for the 1D and
2D GPEs with the respective potential boxes is confirmed by the comparison
of numerical simulations of the underlying GPE and the corresponding amended
COM equation of motion. The reported findings are relevant to the ongoing
experiments carried out for atomic Bose-Einstein condensates trapped in the
box potentials.
\end{abstract}

\maketitle

\section{Introduction}

The Ehrenfest theorem (ET), known since 1927 \cite{Ehrenfest,Ruark}, is the
fundamental tenet of quantum mechanics, which states that the center of mass
(COM) of a wave packet moves according to the respective classical Newton's
equation, in which the role of the potential force is played by its
quantum-mechanical expectation value \cite{LL,B1,B2}. On the other hand, it
was found that the ET in its canonical form does not apply to the
one-dimensional (1D) quantum-mechanical setting including an infinitely deep
potential box \cite{1,2,3,4} (see also a general discussion of the
applicability of the ET in Ref. \cite{5}). Indeed, the formal application of
the ET to the particle in the box obviously predicts that the COM of any
wave packet must perform shuttle motion with a constant velocity, as no
force acts upon the particle inside the box. However, this is not true, as
the analytical solution of the corresponding Schr\"{o}dinger equation
readily produces exact wave functions which feature time-periodic motion of
the COM with variable acceleration (see, e.g., Eq. (\ref{Ehr2}) and Fig. \ref%
{f12}(b) below). A qualitative explanation to this effect, which, in
principle, may be considered as an example of the phenomenology of
quantum anomalies" \cite{an1,an2,an3,an4,Fujikawa,an5,we},
is provided by effective forces exerted over the quantum particle by the
walls (edges) of the 1D potential box. Taking these forces into account
makes it possible to amend the ET, validating it for the quantum particle
trapped in the 1D potential box \cite{3,4},

To the best of our knowledge, the modification of the ET for quantum
particles in a two-dimensional (2D) infinitely deep potential box of the
square shape was not previously reported. This setup is made experimentally
relevant by the realization of optical boxes trapping atomic Bose-Einstein
condensates (BECs) \cite{well} \cite{well-early,well1,well2,well} (see also
a theoretical work \cite{Sadhan}). Furthermore, the experimentally reported
trapping of atomic Bose-Einstein condensates suggests to extend the ET for
the nonlinear Gross-Pitaevskii equation (GPE), instead of the linear
single-particle Schr\"{o}dinger equation. The subject of the present work is
the derivation of the extended (\textit{amended}) ET\ for the 2D and 1D GPEs
with the self-attractive or repulsive cubic nonlinearity, which are subject
to the zero (Dirichlet) boundary conditions (BC) representing the 2D or 1D
potential boxes. The amended ET for the 2D and 1D versions of the GPE is
derived in Section 2. Basic analytical and numerical solutions, which
corroborate and illustrate the amended ET in both the 2D and 1D cases, are
produced, in a concentrated form, in Section 3. The paper is concluded by
Section 4.

\section{The amended ET for the 2D and 1D potential boxes}

The 2D GPE for the mean-field wave function $\psi \left( x,y,t\right) $ is
written, in the scaled notation, as
\begin{equation}
i\frac{\partial \psi }{\partial t}=-\frac{1}{2}\nabla ^{2}\psi +U(x,y)\psi
+g|\psi |^{2}\psi  \label{GP}
\end{equation}%
\cite{an2}, where $U\left( x,y\right) $ is the trapping potential, while $g=-1$ or $+1$ corresponds to the self-attraction or repulsion, respectively.
When $g=0$, Eq. (\ref{GP}) reduces to the linear Schr\"{o}dinger equation.
The infinitely deep potential box is represented by the Dirichlet BC at
edges of the square-shaped area (rectangular box) of size $L$, i.e.,%
\begin{equation}
\psi \left( x=\pm L/2,|y|\leq L/2\right) =\psi \left( y=\pm L/2,|x|\leq
L/2\right) =0,  \label{Diri}
\end{equation}%
while the smooth potential is absent inside the box, i.e. $U=0$. If these BC
are adopted, the solution to Eq. (\ref{GP}) is considered only inside the
square. The 1D version of the model is produced by the reduction of Eqs. (%
\ref{GP}) and (\ref{Diri}) to the single coordinate $x$, as shown below.

To derive the ET, we define the COM coordinates, $\mathbf{R}=\left(
X,Y\right) $, as
\begin{equation}
X=\frac{1}{N}\int \int x\left\vert \psi \left( x,y\right) \right\vert
^{2}dxdy,\;Y=\frac{1}{N}\int \int y\left\vert \psi \left( x,y\right)
\right\vert ^{2}dxdy  \label{R}
\end{equation}%
where $N=\int \int |\psi |^{2}dxdy$ is the norm of the mean-field wave
function, which is proportional to the number atoms in BEC. If the GPE is
subject to the zero BC (\ref{Diri}), the integration in Eq. (\ref{R}) is
constrained to the inner area of the potential box.

First, we derive an expression for the COM velocity, $d\mathbf{R}/dt$,
differentiating expression (\ref{R}) and substituting $\partial \psi
/\partial t$ as per Eq. (\ref{GP}):
\begin{equation}
\frac{d\mathbf{R}}{dt}=\frac{1}{N}\int \int \mathbf{r}\left( \frac{i}{2}\psi
^{\ast }\nabla ^{2}\psi +\mathrm{c.c.}\right) dxdy,  \label{dR/dt}
\end{equation}%
where $\mathbf{r}=\left( x,y\right) $, with both $\ast $ and $\mathrm{c.c.}$%
\ standing for the complex conjugate (the terms $\sim U$ and $\sim g$
immediately cancel out in Eq. (\ref{dR/dt})). Further, expression (\ref%
{dR/dt}) can be transformed by means of the integration by parts, using the
following identity in the integrand:%
\begin{equation}
ix_{j}\psi ^{\ast }\partial _{kk}^{2}\psi +\mathrm{c.c.}\equiv \left[
i\partial _{k}\left( x_{j}\psi ^{\ast }\partial _{k}\psi \right)
-ix_{j}\left\vert \partial _{k}\psi \right\vert ^{2}-i\psi ^{\ast }\partial
_{j}\psi \right] +\mathrm{c.c.},  \label{identity}
\end{equation}%
where $j,k=1,2$,$~x_{1}\equiv x$, $x_{2}\equiv y$, and the Einstein's
convention is adopted, i.e., automatic summation over repeating indices ($k$%
, in this case). If identity (\ref{identity}) is substituted in Eq. (\ref%
{dR/dt}), the integral of the first (full-derivative) term vanishes, in both
cases of the infinite area and the one constrained by BC (\ref{Diri}) (in
the latter case, the result of the integration of the full derivative, which
amounts to the \textit{surface term} taken at the edge of the square, is
zero pursuant to BC (\ref{Diri})). The substitution of the second term from
expression (\ref{identity}) in Eq. (\ref{dR/dt}) yields zero too, as this
term is imaginary. Thus, Eq. (\ref{dR/dt}) is reduced to the form with the
right-hand side (RHS) determined by the third term from the RHS of Eq.
(identity):
\begin{equation}
\frac{d\mathbf{R}}{dt}=\frac{i}{2N}\int \int \psi \nabla \psi ^{\ast }dxdy+%
\mathrm{c.c.}  \label{dR/dt 2}
\end{equation}

To continue the derivation of the (amended) ET, expression (\ref{dR/dt 2})
is repeatedly differentiated with respect to $t$, again substituting $%
\partial \psi /\partial t$ as per Eq. (\ref{GP}). After straightforward
manipulations similar to those outlined above (which include the integration
by parts and dropping the surface terms which are nullified by BC (\ref{Diri}%
)), an intermediate result is
\begin{equation}
\frac{d^{2}\mathbf{R}}{dt^{2}}=\frac{1}{N}\int \int \left[ -\frac{1}{2}%
\nabla ^{2}\psi +U(r)\psi +g|\psi |^{2}\psi \right] \nabla \psi ^{\ast }dxdy+%
\mathrm{c.c.}  \label{second}
\end{equation}%
Further consideration easily shows that the term $\sim g$ in Eq. (\ref%
{second}) yields zero contribution, the term $\sim U$ yields, with the help
of the integration by parts, precisely the quantum-mean ($\mathrm{qmean}$,
alias expectation) value of the potential force,
\begin{equation}
-\left( \nabla U(R)\right) _{\mathrm{qmean}}\equiv -\frac{1}{N}\int \int
\nabla U(r)\left\vert \psi \left( x,y\right) \right\vert ^{2}dxdy,
\label{qmean}
\end{equation}
and there remains an additional \textquotedblleft anomalous"
integro-gradient term in the resulting expression for the COM's
acceleration, which originates from the quantum-pressure term $-\frac{1}{2}%
\nabla ^{2}\psi $ in the integrand on the RHS of Eq. (\ref{second}):
\begin{equation}
\frac{d^{2}\mathbf{R}}{dt^{2}}=-\left( \nabla U(R)\right) _{\mathrm{qmean}}-%
\frac{1}{2N}\left[ \int \int \nabla ^{2}\psi ^{\ast }\cdot \nabla \psi dxdy+%
\mathrm{c.c.}\right] .  \label{second 2}
\end{equation}
Writing the integrand of the second term on the right-hand side of Eq. (\ref%
{second 2}) with the explicit coordinate indices (cf. Eq. (\ref{identity})),
it is easy to represent it in the form of a full derivative:
\begin{equation}
-\frac{1}{2N}\left( \partial _{j}\psi \cdot \partial _{kk}^{2}\psi ^{\ast
}+\partial _{j}\psi ^{\ast }\cdot \partial _{kk}^{2}\psi \right) \equiv
\frac{1}{2N}\left[ -\partial _{k}\left( \partial _{j}\psi \cdot \partial
_{k}\psi ^{\ast }+\partial _{j}\psi ^{\ast }\cdot \partial _{k}\psi \right)
+\partial _{j}\left( \partial _{k}\psi ^{\ast }\cdot \partial _{k}\psi
\right) \right] ,  \label{identity 2}
\end{equation}%
In the case of a localized wave function in the infinite area, the integral
of the full-derivative term (\ref{identity 2}) immediately vanishes,
reducing Eq.~(\ref{second 2}) to the usual ET\ form:
\begin{equation}
\frac{d^{2}\mathbf{R}}{dt^{2}}=-\left( \nabla U(R)\right) _{\mathrm{qmean}}.
\end{equation}%
In particular, if $\psi (x,y)$ is highly localized, $(\nabla U(R))_{\mathrm{%
qmean}}$ may be approximated at $\nabla U(R)$, producing the Newton's
equation of motion from classical mechanics.

A well-known example is Eq. (\ref{GP}) in the infinite area with the
harmonic-oscillator (HO) trapping potential (in the scaled form),
\begin{equation}
U_{\mathrm{HO}}=(1/2)(x^{2}+y^{2}).  \label{HO}
\end{equation}
In that case, the last term in Eq. (\ref{second 2})
vanishes, as said above, while the remaining form of Eq. (\ref{second 2}) is
fully identical to the classical equation of motion,
\begin{equation}
\frac{d^{2}\mathbf{R}}{dt^{2}}=-\mathbf{R}  \label{RR}
\end{equation}
because
\[-\frac{1}{N}\int\int \nabla U(r)|\psi(x,y)|^2dxdy=-\frac{1}{N}\int\int\mathbf{r}|\psi(x,y)|^2dxdy=-\mathbf{R}.\]

On the other hand, in the case of the infinitely deep box with $U=0$,
defined by BC (\ref{Diri}), the entire effective force in the COM equation
of motion (\ref{second 2}) is produced by the \textquotedblleft anomalous"
term, which may be construed as the result of the interaction of the trapped
condensate with the edges of the potential box. Indeed, the integration of
the full-derivative term in Eq. (\ref{second 2}), with $U=0$ (no potential
acting inside the box) casts the COM\ equation of motion in the following
form, with rather cumbersome expressions for the components of the
edge-exerted force:%
\begin{eqnarray}
\frac{d^{2}X}{dt^{2}} &=&\frac{1}{N}\left[ \frac{1}{2}\int_{-L/2}^{+L/2}%
\left( \left\vert \partial _{y}\psi \right\vert ^{2}-\left\vert \partial
_{x}\psi \right\vert ^{2}\right) dy{\LARGE |}_{x=-L/2}^{x=+L/2}-%
\int_{-L/2}^{+L/2}\mathrm{Re}\left( \partial _{x}\psi \cdot \partial
_{y}\psi ^{\ast }\right) dx{\LARGE |}_{y=-L/2}^{y=+L/2}\right] \equiv F_{x},
\label{X} \\
\frac{d^{2}Y}{dt^{2}} &=&\frac{1}{N}\left[ \frac{1}{2}\int_{-L/2}^{+L/2}%
\left( \left\vert \partial _{x}\psi \right\vert ^{2}-\left\vert \partial
_{y}\psi \right\vert ^{2}\right) dx{\LARGE |}_{y=-L/2}^{y=+L/2}-%
\int_{-L/2}^{+L/2}\mathrm{Re}\left( \partial _{y}\psi \cdot \partial
_{x}\psi ^{\ast }\right) dy{\LARGE |}_{x=-L/2}^{x=+L/2}\right] \equiv F_{y},
\label{Y}
\end{eqnarray}%
where ${\LARGE |}_{x,y=-L/2}^{x,y=+L/2}$ implies, as usual, taking the
difference of the expression at points $x,y=\pm L/2$. Similar expressions
can be derived for 2D potential boxes of different shapes, such as a
cylindrical one.

In the case of the 1D infinitely deep box, the integration in the 1D version
of Eq. (\ref{second 2}) (with $U=0$) can be easily performed, leading to the
following COM\ equation of motion:
\begin{equation}
\frac{d^{2}X}{dt^{2}}=\frac{1}{2N}\left\vert \partial _{x}\psi \right\vert
^{2}{\LARGE |}_{x=-L/2}^{x=+L/2}~\equiv F_{\mathrm{1D}}.  \label{Ehr2}
\end{equation}%
On the other hand, the \textquotedblleft anomalous" force does not appear in
the 1D ring-shaped setup, which is modeled by the 1D GPE subject to the
periodic BC.

A relevant generalization of the underlying GPE (\ref{GP}) is the one in
which the nonlinearity coefficient $g$ may be spatially dependent, $g\left(
x,y\right) $ \cite{WMLiu}. Because $g$ does not explicitly appear in Eqs. (%
\ref{X}), (\ref{Y}) and (\ref{Ehr2}), these equations remain valid in this
case too.

\section{Numerical results}

\subsection{The 2D system}

First, we provide typical examples of the COM motion in the framework of the
2D GPE (\ref{GP}) which includes the HO trapping potential (\ref{HO}) and
the self-defocusing, zero, or focusing nonlinearity, corresponding to $%
g=+1,0 $ or $-1$, respectively. The initial condition is taken as a linear
combination of the ground state and lowest vortex eigenmode of the 2D linear
HO (the one with $g=0$) \cite{LL}, with the ratio of the real participation
coefficients $1:a$, i.e.,%
\begin{equation}
\psi _{0}(x,y)=\aleph \exp \left( -\frac{1}{2}\left( x^{2}+y^{2}\right)
\right) \left[ 1+a(x+iy)\right] ,  \label{0}
\end{equation}%
where $\aleph $ is a real amplitude. The 2D norm of input (\ref{0}) is
\begin{equation}
N=\pi \aleph ^{2}\left( 1+a^{2}\right) ,  \label{N}
\end{equation}

At point
\begin{equation}
x=x_{p}\equiv \frac{\sqrt{1+4a^{2}}-1}{2a},~y=0,  \label{xp}
\end{equation}
the input (\ref{0}) attains its maximum (peak) value,%
\begin{equation}
A_{p}=\aleph \exp \left( -x_{p}^{2}/2\right) \left( 1+ax_{p}\right) .
\label{Ap}
\end{equation}%
The COM coordinates (\ref{R}) of input (\ref{0}) are
\begin{equation}
X=a/\left( 1+a^{2}\right) \equiv x_{\mathrm{COM}},Y=0.  \label{COM}
\end{equation}%
Obviously, the linear Schr\"{o}dinger equation with input (\ref{0}) produces
an elementary solution for the wave function,%
\begin{equation}
\psi (x,y;t)=\aleph \exp \left( -\frac{1}{2}\left( x^{2}+y^{2}\right)
\right) \left[ e^{-it}+a(x+iy)e^{-2it}\right] .  \label{psi}
\end{equation}%
In the course of the evolution governed by this solution, the peak-value
point moves with angular frequency $\omega =1$ along the ring $\sqrt{%
x^{2}+y^{2}}=x_{p}$, keeping the constant peak value (\ref{Ap}) of $%
\left\vert \psi \left( x,y,t\right) \right\vert $.

\begin{figure}[h]
\begin{center}
\includegraphics[height=3.5cm]{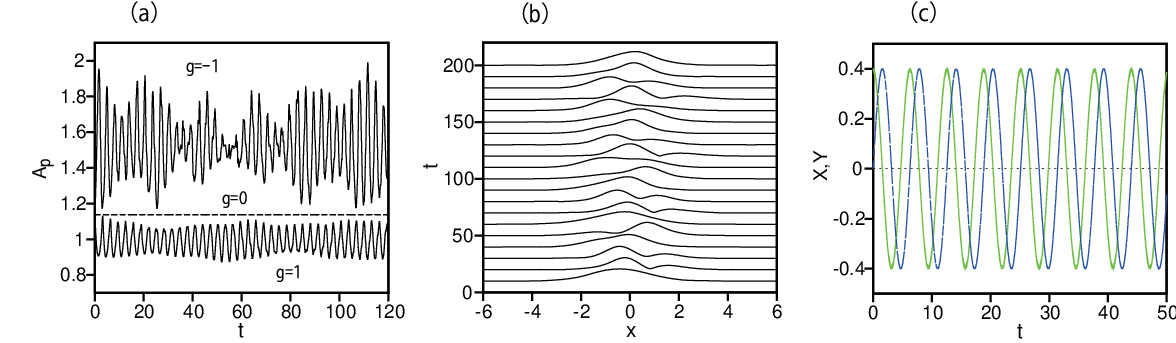}
\end{center}
\caption{(a) The evolution of the peak (largest) value of $|\protect\psi %
(x,y,t)|$ for $g=-1$ (the upper solid line), $g=0$ (the flat dashed line),
and $g=+1$ (the lower solid line), as produced by the numerical simulations
of Eq. (\protect\ref{GP}) with the HO potential (\protect\ref{HO}) and input
(\protect\ref{0}). (b) The evolution of $|\protect\psi (x,y,t)|$ in the
cross section $y=0$ for $g=-1$. (c) The evolution of the COM coordinates $X$
and $Y$ (the solid green and dashed blue lines, respectively), as produced
by the data of the numerical simulations, as per definitions (\protect\ref{R}%
) for $g=+1$, $0$, and $-1$. The ET-predicted dependences (\protect\ref{XY})
of the COM\ coordinates are plotted by the solid green and dashed blue
lines, respectively. The parameters of the input (\protect\ref{0}) are $%
N=5.7 $ and $a=2$, hence the other parameters are $\aleph \approx 0.60$, $%
A_{p}\approx 1.14$, and $x_{\mathrm{COM}}=0.4$, see Eqs. (\protect\ref{N}), (%
\protect\ref{Ap}), and (\protect\ref{COM}).}
\label{f10}
\end{figure}

Characteristic examples of the evolution of the input (\ref{0}) for the
three above-mentioned cases, $g=+1$, $0$, and $-1$, are displayed in Fig. %
\ref{f10}, for the fixed values of the norm, $N=5.7$, and $a=2$ in Eq. (\ref%
{0}) (hence, the amplitude is $\aleph \approx 0.60$, and the peak value is $%
A_{p}\approx 1.14$, according to Eqs. (\ref{xp}) and (\ref{Ap})). The
simulations were performed in the square of size $L\times L=12\times 12$, by
means of the split-step Fourier method based on a set of $256\times 256$
modes.

In full agreement with the\ normal ET, it is observed that, while the
evolution of the wave function as a whole is quite different in the cases of
the self-repulsion, no self-interaction, and self-attraction (as specified,
in particular, by the evolution of the peak value displayed in Fig. \ref{f10}%
(a)), the evolution of the COM coordinates $X(t)$ and $Y(t)$ is identical in
all the cases, given by an obvious solution of Eq. (\ref{RR}):%
\begin{equation}
X(t)=x_{\mathrm{COM}}\cos t,~Y(t)=x_{\mathrm{COM}}\sin t,  \label{XY}
\end{equation}%
with $x_{\mathrm{COM}}$\ defined by Eq. (\ref{COM}). The corresponding COM\
trajectory in the $\left( x,y\right) $ plane is the ring of radius $x_{%
\mathrm{COM}}$.

Note that the 2D GPE with $g=-1$ may give rise to the wave collapse
(catastrophic self-compression of the wave function) \cite{Berge,Kuz,Fibich}
if the input's norm exceeds the well-known critical value, $N_{\mathrm{cr}%
}\approx 5.85$ (in the present notation; its approximate variational value
is $\left( N_{\mathrm{cr}}\right) _{\mathrm{var}}=2\pi $ \cite{Anderson}).
The proximity of $N=5.7$ to $N_{\mathrm{cr}}$ leads to irregular
large-amplitude oscillations of $A_{p}(t)$ in Fig. \ref{f10}(a) and
irregular spatiotemporal evolution in Fig. \ref{f10}(b); nevertheless, this
irregularity produces no effect on the ET-supported strictly harmonic
oscillations of the COM coordinates in Fig. \ref{f10}(c). Furthermore,
additional simulations of the collapse development in the case of the input
with $N>N_{\mathrm{cr}}$ (not shown here in detail) corroborate that the
strictly harmonic oscillations of $X(t)$ and $Y(t)$ persist even in this
case, as long as the appearance of very large amplitudes, generated by the
collapse, does not break the simulation algorithm.

\begin{figure}[h]
\begin{center}
\includegraphics[height=6.5cm]{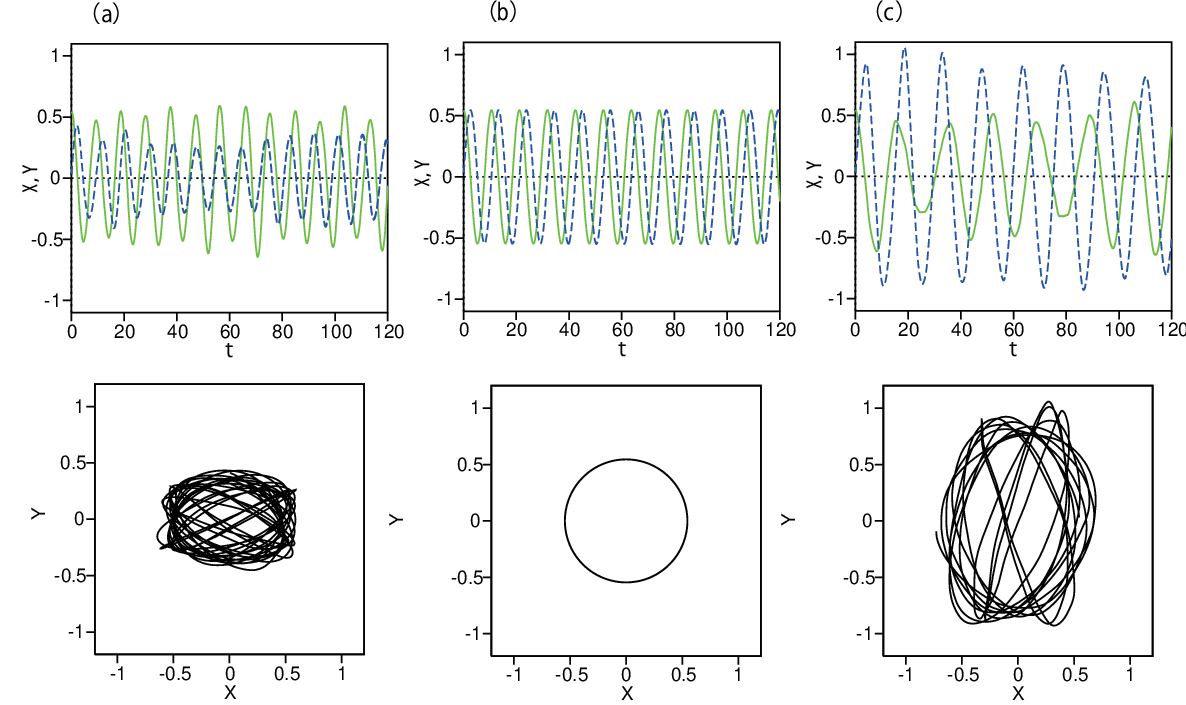}
\end{center}
\caption{The evolution of the COM coordinates $X(t)$ and $Y(t)$ (the solid
green and dashed blue lines, respectively), as directly produced by the
simulations of the 2D GPE (\protect\ref{GP}) with $U=0$, BC (\protect\ref%
{Diri}), and input (\protect\ref{2D0}). These dependences are
indistingushable from their counterparts produced by the equations of motion
(\protect\ref{X}) and (\protect\ref{Y}), which represent the amended ET for
the 2D potential box. Panels (a), (b), and (c) represent, severally, the GPE
with $g=+1$, $g=0$, and $g=-1$. The corresponding COM\ trajectories in the $\left(
x,y\right) $ plane are plotted in bottom panels. The parameters are $N=5$, $%
a=0.4$, and $L=10$.}
\label{f14}
\end{figure}

The central point of the present work is to consider the dynamics governed
by the GPE in the 2D potential box with $U=0$, subject to BC (\ref{Diri}),
for which the ET takes the amended form given by Eqs. (\ref{second 2}) or (%
\ref{X}) and (\ref{Y}). A set of characteristic examples is displayed in
Fig. \ref{f14}, taking the initial condition as a superposition of obvious
eigenstates of the 2D linear Schr\"{o}dinger equation with BC (\ref{Diri}),
\begin{equation}
\psi _{0}(x,y)=\aleph \{\cos (kx)\cos (ky)+a\left[ \sin (2kx)\cos (ky)+i\cos
(kx)\sin (2ky)\right] \},  \label{2D0}
\end{equation}%
where $k=\pi /L$, and the amplitude is matched to a given value of the 2D
norm, $\aleph =\sqrt{4N/\left[ L^{2}(1+2a^{2})\right] }$. This input is
roughly similar to the above one (\ref{0}), with the first term representing
the ground state of the linear system, and the second one, with the real
relative-participation amplitude $a$, emulating the vortex term in Eq. (\ref%
{0}). Characteristic examples, produced by the numerical simulations of Eq. (%
\ref{GP}) with $U=0$, BC (\ref{Diri}), and input (\ref{2D0}), are presented
in Fig. \ref{f14} for norm $N=5$ and relative participation amplitude $a=0.4$
in expression (\ref{2D0}) (in particular, the dynamics displayed in panel
(c) does not lead to the collapse in the case of the self-focusing, as $N=5$
is smaller than the above-mentioned critical value, $N_{\mathrm{cr}}\approx
5.85$).

The solution of the linear version of Eq. (\ref{GP}), with $U=g=0$ and BC (%
\ref{Diri}), produced by input (\ref{2D0}), is obvious:
\begin{equation}
\psi (x,y,t)=\aleph \left\{ \cos (kx)\cos (ky)e^{-i\omega _{1}t}+a\left[
\sin (2kx)\cos (ky)e^{-i\omega _{2}t}+ia\cos (kx)\sin (2ky)\right]
e^{-i\omega _{2}t}\right\} ,  \label{psi2D}
\end{equation}%
with eigenfrequencies $\omega _{1}=k^{2}$ and $\omega _{2}=5k^{2}/2$. In
this case, the components of the effective edge-exerted force in the 2D
potential box, defined by the second term on the right-hand side of Eq. (\ref%
{second 2}), or, alternatively, by the more explicit expressions (\ref{X})
and (\ref{Y}), are readily calculated as
\begin{equation}
F_{x}=-\frac{8\pi ^{2}a}{L^{3}\left( 1+2a^{2}\right) }\cos \left( (\omega
_{2}-\omega _{1})t\right) ,\;F_{y}=-\frac{8\pi ^{2}a}{L^{3}\left(
1+2a^{2}\right) }\sin \left( (\omega _{2}-\omega _{1})t\right) .  \label{ff2}
\end{equation}%
In the framework of the nonlinear GPE, with $g=\pm 1$, it is necessary to
obtain numerical solutions of Eq. (\ref{GP}) and then compute the effective
force, substituting the numerical solutions in Eq. (\ref{second 2}) or Eqs. (%
\ref{X}) and (\ref{Y}).

For the GPE in the 2D potential box with the self-defocusing ($g=+1$) and
focusing ($g=-1$) nonlinearities, as well as for the linear GPE (with $g=0$%
), Figs. \ref{f14}(a), (c), and (b) display, severally, the evolution of the
COM coordinates $X$ and $Y$ (the green solid and blue dashed lines,
respectively), as directly obtained from the numerical simulations. The $X(t)
$ and $Y(t)$ dependences are indistinguishable from their counterparts
produced by the amended ET, i.e., Eq. (\ref{second 2}) or Eqs. (\ref{X}) and
(\ref{Y}), thus corroborating the amended ET for the 2D potential box in
both the linear and nonlinear cases. Note that, unlike the usual ET (cf.
Fig. \ref{f10}), the amended version predicts irregular (apparently chaotic)
COM\ motion in the strongly nonlinear setup. The chaotic character of the
motion under the action of the nonlinearity ($g\neq 0$) is corroborated by
positive values of the respective Lyapunov exponents. The largest exponents
are numerically evaluated as $2.0\times 10^{-3}$ at $g=+1$, $-1.5\times
10^{-7}$ at $g=0$, and $1.9\times 10^{-3}$ at $g=-1$.

\subsection{Reduction to the 1D potential box}

The underlying 2D GPE (\ref{GP}) with $U=0$, subject to BC (\ref{Diri}), can
be reduced to the 1D form by considering a narrow potential box, of size $%
(L,l)$ in the $x$ and $y$ directions, with $l\ll L$, taking the limit of $%
l\rightarrow 0$, and performing integration in the $y$ direction. Thus one
arrives at the 1D version of Eq. (\ref{second 2}),
\begin{equation}
i\frac{\partial \psi }{\partial t}=-\frac{1}{2}\frac{\partial ^{2}\psi }{%
\partial x^{2}}+g|\psi |^{2}\psi ,  \label{1D}
\end{equation}%
with BC $\psi (x=\pm L/2)=0$ and the nonlinearity coefficient set, as above,
to be $g=\pm 1$ or $0$. In this connection, it is relevant to mention
analytical stationary equations of Eq. (\ref{1D}) with this BC, which were
found for $g=+1$ \cite{Carr1} and $g=-1$ \cite{Carr2}.

Here we display the results for the fixed norm $N=\int_{-L/2}^{+L/2}|\psi
(x)|^{2}dx=5$ and the 1D-box size $L=4$. The initial condition is chosen as
a mixture of the ground and first excited states, roughly similar to the 2D
inputs (\ref{0}) and (\ref{2D0}):
\begin{equation}
\psi _{0}(x)=\aleph \left[ \cos (\pi x/L)+a\sin (2\pi x/L)\right] ,
\label{1D0}
\end{equation}%
with the amplitude expressed in terms of the norm as $\aleph =\sqrt{2N/\left[
L(1+a^{2})\right] }$. In the case of $g=0$, the solution to Eq. (\ref{1D})
with input (\ref{1D0}) is obvious:
\begin{equation*}
\psi (x,t)=\aleph \left[ \cos (\pi x/L)e^{-i\omega t}+a\sin (2\pi
x/L)e^{-4i\omega }\right] ,
\end{equation*}%
with $\omega =\pi ^{2}/\left( 2L^{2}\right) $, the corresponding effective
edge-generated force (\ref{Ehr2}) being%
\begin{equation}
F_{\mathrm{1D}}=-\frac{8\pi ^{2}a}{L^{3}(1+a^{2})}\cos \left( 3\omega
t\right) ,  \label{ff}
\end{equation}%
cf. Eq. (\ref{ff2}).

\begin{figure}[h]
\begin{center}
\includegraphics[height=3.5cm]{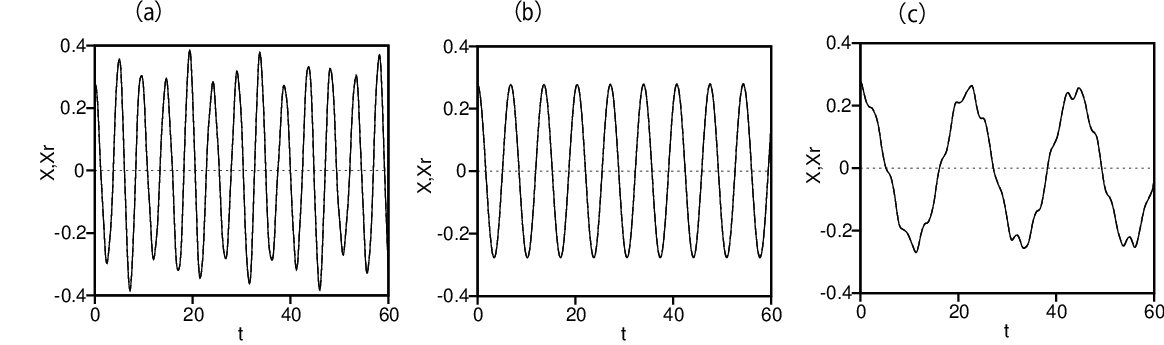}
\end{center}
\caption{The evolution of the COM coordinate $X$ (the solid line), as
produced by the direct simulations of Eq. (\protect\ref{1D}) with input (%
\protect\ref{1D0}) in the 1D potential box, and its counterpart $X_{r}$ (the
dashed line), as reconstructed by means of the amended-ET equation (\protect
\ref{Ehr2}), for $g=+1$ (a), $g=0$ (b), and $g=-1$ (c) .}
\label{f12}
\end{figure}

The results are given here for the relative participation coefficient of the
first excited state $a=0.2$ in Eq. (\ref{1D0}), hence $\aleph \approx 1.57$.
The simulations were performed by means of the split-step Fourier method
with $512$ modes. Solid curves in Figs. \ref{f12}(a), (b), and (c) display
the evolution of the COM coordinate, $X(t)=N^{-1}\int_{-L/2}^{+L/2}x|\psi
\left( x,t\right) |^{2}dx$ (solid lines), for $g=+1$, $0$, and $-1$,
respectively, as directly computed from the data of the simulations of Eq. (%
\ref{1D}). In the same figures, the dashed curves, which are actually
indistinguishable from the solid ones, represent the COM coordinate, $%
X_{r}(t)$, as obtained from Eq. (\ref{Ehr2}). Thus, the amended ET is fully
corroborated in the 1D case too.

\section{Conclusion}

The usual form of the ET (Ehrenfest theorem), which was established about
100 years ago \cite{Ehrenfest}, states that the motion of the COM (center of
mass) of the 1D quantum-mechanical wave packet is governed by the
corresponding classical equation of motion, in which the usual force is
replaced by its expectation quantum-mechanical value (it is identical to the
classical force in the case of the HO (harmonic-oscillator) potential). More
recently, it has been found that ET is not valid, in this form, for the
quantum-mechanical particle trapped in the infinitely deep 1D potential box.
An alternative form of the ET was derived for this case, with an effective
force induced by the interaction of the particle with the edges of the 1D
box. In the present work, we have extended the concept of the amended ET for
the linear Schr\"{o}dinger equation replaced by the nonlinear GPE
(Gross-Pitaevskii equation) with the cubic self-focusing or defocusing term,
as well as for the 2D potential box. In the latter case, we have derived an
effective force induced by edges of the square-shaped box, while the
nonlinearity does not add a direct contribution to the ET, in the 1D and 2D
cases alike. Nevertheless, the nonlinearity affects the amended ET through
the edge-generated force. In particular, the presence of the nonlinear terms
in the underlying GPE can make the COM motion in the potential box irregular
(chaotic). The validity of the amended ET in both the 1D and 2D settings is
fully corroborated by the comparison with results of the full numerical
simulations. These results are directly relevant to the current experiments
with atomic BEC trapped in rectangular potential boxes.

As an extension of the analysis, it may be interesting to develop it for a
system of nonlinearly coupled GPEs modeling a binary BEC \cite{WMLiu2}. Such
a generalization may be straightforward, as the above-mentioned equations (%
\ref{X}), (\ref{Y}) and (\ref{Ehr2}) will not include terms accounting for
the interaction between the two components, hence the equations will keep
their form separately for each component. Another relevant possibility is to
consider a 2D cylindrical potential box, as well as a 3D one of the cubic
form, as the trapping potentials of these types are also available to the
experiment. On the other hand, a challenging possibility may be to introduce
an amended ET in relativistic quantum mechanics, cf. Refs. \cite%
{Dirac1,BBBB,Dirac2}.

\end{document}